\documentclass{elsart}

 \usepackage{graphics}
 \usepackage{graphicx}
 \usepackage{epsfig}

\usepackage{amssymb}

\begin{document}

\begin{frontmatter}



\title{Inter-package dependency networks in open-source software}


\author[UNI]{Nathan LaBelle\corauthref{cor1}}
\author[UNI]{Eugene Wallingford}

\corauth[cor1]{Corresponding author}

\address[UNI]{Computer Science Department, University of Northern Iowa, Cedar Falls, Iowa 50613}

\begin{abstract}
This research analyzes complex networks in open-source software at the inter-package level, where package 
dependencies often span across projects and between development groups. We review complex networks identified at 
``lower'' levels of abstraction, and then formulate a description of interacting software components at the package 
level, a relatively ``high'' level of abstraction. By mining open-source software repositories from two sources, we 
empirically show that the coupling of modules at 
this granularity creates a small-world and scale-free network in both instances. 
\end{abstract}

\end{frontmatter}

\section{Introduction and Previous Research}
\label{}

\par In recent years the identification and categorization of networks has become an emerging research area in fields 
as diverse as sociology and biology, but has remained relatively unutilized in software engineering. The study and categorization of software systems as networks is a promising field, as the the identification of 
networks in software systems may prove to be a valuable tool in managing the complexity and dynamics of software 
growth, which have traditionally been problems in software engineering. However, current trends in software 
development offer diverse 
and accessible software to study, which may help software engineers learn how to create better programs.  In 
particular, open-source software (OSS) allows researchers access to a rich set 
of examples that are production-quality and studiable ``in the wild''. They are a valuable asset that 
can aid in the 
study of software development and managing complexity. 
\par In OSS systems, applications are often distributed in the form of packages. A package is a bundle 
of related components necessary to compile or run an application. Because resource reuse is naturally a pillar of 
OSS, a package is often dependent on some other packages to function properly. These packages may be third-party 
libraries, bundles of resources such as images, or Unix utilities such as grep and sed. Package dependencies often 
span 
across project development teams, and since there is no central control over which resources from other packages are needed, the software system 
self-organizes in to a collection of discrete, interconnected components. This research applies complex network 
theory to package dependency networks mined from two OSS repositories. 
\par A network is a large (typically unweighted and simple) graph $ G=(V,E) $ where $ V $ denotes a vertex 
set and $ E $ an edge set. Vertices represent discrete objects in a dynamical system, such as social actors, economic agents, computer programs, or biological 
producers and consumers. Edges represent interactions among these ``interactons''. For example, if software objects 
are represented as vertices, edges can be assembled between them by defining some meaningful interaction between 
the objects, such as inheritence or procedure calls (depending on the nature of the programming language used).
\par Real-world networks tend to share a common set of non-trivial properties: they have scale-free degree 
distributions and exhibit the small-world effect. The degree of a vertex $v$, denoted $k$, is the number of 
vertices adjacent to $v$, or in the case of a digraph  either the number of incoming edges or outgoing edges, 
denoted $k_{in}$ and $k_{out}$, respectively. In real-world networks such as the Internet \cite{FFF99}, the 
World-Wide Web \cite{AJB99}, software objects \cite{CMy03,PNF04,WC03,VS04}, networks of scientic citations 
\cite{LLJ03,SRe98}, the distribution of edges roughly follows a power-law: $P(k) 
\propto k^{ - a} $. That is, the probability of a vertex having $k$ edges decays with respect to some 
constant $a \in \Rset^+$. This is significant because it shows deviation from randomly constructed graphs, first 
studied by Erd$\ddot{o}$s and R$\acute{e}$nyi and proven to take on a Poisson distribution in the limit of large $n$, where 
$n=|V|$ \cite{BBo85}. 
\par Random connection models also fail to explain the ``small-world effect'' in real networks, the canonical 
examples being social collaboration networks \cite{MNe00,MNe04}, 
certain neural networks \cite{WS98}, and the World-Wide Web \cite{AJB99}.  The small-world effect states that 
$C_{random} \ll C_{sw}$ and $L_{random}\approx L_{sw}$ where $C$ is the {\em clustering coefficient} of a graph, 
and $L$ is the s {\em characteristic path length} \cite{WS98}. The clustering coefficent is the propensity for 
neighbors $u,w \in V$ of a vertex $v$ to be connected to each other. For a vertex $v$, we can define the clustering coefficent as 
$C_v=\frac{k_v}{{n \choose 2}}$, and therefore $C_v \in [0,1]$. The clustering coefficient for a graph is the average 
over all vertices, $C=\frac{1}{n}\sum_{v \in V}{C_v}$. Real-world networks are normally highly clustered while random networks are not, 
because $C_{random} = \frac{\bar{k}}{n}$ for large networks \cite{BBo85}. Because most networks are sparse, that is 
$n \gg k$, random networks are not highly clustered. $L$ is the average geodesic (unweighted) distance between vertices. 
\par To summarize, random graphs are not small-world because they are not highly clustered (although they have 
short path lengths) and they are do not follow the commonly observed power-law  because the edge distribution is Poissonian. The presence of 
these features in networks indicate non-random creation mechanisms, which although several models have been proposed, 
none is agreed upons. In order to make accurate hypothesis about possible network creation mechanisms, a wide variety 
of real-world networks sharing these non-trivial properties should be identified.
\par Previous research in networks of software have focused on software at``low'' levels of abstraction (relative 
to the 
current research). Clark and Green \cite{CG77} found Zipf distributions (a ranking distribution similar to the 
power-law, which is also found in word frequencies in natural language \cite{Zipf}) in the structure of CDR and CAR 
lists in large Lisp programs during run-time. In the case of object-oriented programming languages, several studies 
\cite{CMy03,PNF04,VS04,WC03} have 
identified the small-world effect and power-law edge distribution in networks of objects or procedures where edges 
represent
meaningful interconnection between objects, such as inheritence or in the case in procedural languages, 
procedures are represented as vertices and edges between vertices symbolize function calls. Similar statistical features 
have also been identified in networks where the vertices represent source code files on a disk and edges represent a 
dependency between files (for example, in C and C++ one source file may {\em \#include} another) \cite{MLM03}, and in 
documentation systems \cite{WC03}.
\section{Package Dependency Networks} 
\par Mining the Debian GNU/Linux software repository \cite{Deb} and the FreeBSD Ports Collection \cite{BSD} has 
allowed us to create networks of package dependencies. In the case of the Debian repository, data was taken from the i386 
branch of the ``unstable'' tree, which contains the most up-to-date software and is the largest branch. The Debian 
data was extracted using {\em apt} (Advanced Packaging Tool), while the BSD data was extracted from the ports {\em 
INDEX} system. The BSD Ports system allowed us to distinguish between run-time dependencies and 
compile-time (build) dependencies. The data here is for only compile-time dependencies, although results are similar 
for run-time dependencies. Graphs were constructed in Java using the Java Universal Network/Graph framework 
\cite{JUNG}. ``Snapshots'' of the repositories were taken during the month of September, 2004.

The Debian network contains $n=19,504$ packages and $m=73,960$ edges, giving each package an average coupling to 
$3.79$ packages. For the Debian network, $C=0.52$ and $L=3.34$. This puts the Debian network in the small-world 
range, since an equivalent random graph would have $C_{random}\approx.0019$ and $L_{random}\approx7.41$. There are 
1,945 components, but the largest component contains 88\% of the vertices. The rest of the vertices are disjoint 
from each other, resulting in a large number of components with only 1 vertex. The diameter of the largest component 
is 31. The distribution of outgoing edges, which is a measure of dependency to other packages, follows a power-law 
with $\alpha_{out}\approx2.33$. The distribution of incoming edges, which measures how many packages are dependent on 
a package, follows a power-law with $\alpha_{in}\approx0.90$. While 10,142 packages are not referenced by any 
package at all, the most highly referenced packages are referenced thousands of times. 73\% of packages depend on some other 
package to function correctly. Correlation between $k_{in}$, $k_{out}$, and package size is not calculated because 
the normality assumption is violated.

\begin{center}
\begin{tabular}{|l||l|l|}
\hline
& Debian & BSD \\ \hline
$n$ & 19,504 & 10,222 \\ \hline
$m$ & 73,960 & 74,318 \\ \hline
$|\Omega|$ & 17,351 & 7,441 \\ \hline
$\alpha_{in}$ & 0.9 & 0.62 \\ \hline
$\alpha_{out}$ & 2.33 & 1.28 \\ \hline
$C$ & 0.52 &  0.56 \\ \hline
$L$ & 3.34 &  2.86 \\ \hline
\end{tabular}
\end{center}

The BSD compile-time dependency network contains $n=10,222$ packages and $m=74,318$ edges, coupling 
each package to an average of $\bar{k}=7.27$ other packages. For the BSD network, $C\approx0.56$ and $L\approx2.86$. An equivalent random graph 
would have $C_{random}\approx0.007$ and $L_{random}\approx7.11$. Hence, the BSD network is small-world. The degree 
distribution of the BSD network also resembles a power-law, with $\alpha_{in}\approx0.62$ and 
$\alpha_{out}\approx1.28$. For the run-time network, results were similar: the run-time network is both small-world 
and follows a power-law.
\par In the Debian network, the 20 most highly depended-upon packages are libc6 (7861), xlibs (2236), 
libgcc1 (1760), zlib1g (1701), libx11-6 (1446), perl (1356), libxext6 (1110), debconf (1013), libice6 (922), libsm6 (919), 
libglib2.0-0 (859), libpng12-0 (622), libncurses5 (616), libgtk2.0-0 (615), libpango1.0-0 (610), libatk1.0-0 (602), 
libglib1.2 (545), libxml2 (538), libart-2.0-2 (524), and libgtk1.2 (474). The number in parentheses represents the 
number of incoming edges.  The list is composed mainly of libraries that provide some functionality to programs such as XML parsing or that provide some reusable components 
such as graphical interface widgets. Because the most highly-connected package (libc6) is required for execution of C 
and C++ programs, we can infer that these are the  most widely used programming languages.

Figure 1 shows the double-log distribution of edges in the Debian network (scatterplots for the BSD network would 
have a similar shape). From the figure we can see the heavy-tailed power-law shape. The absolute value of the slope 
of the regression line indicates the power-law exponent, $\alpha$.
\begin{figure}
\centering
\epsfysize=1.5in
\epsffile{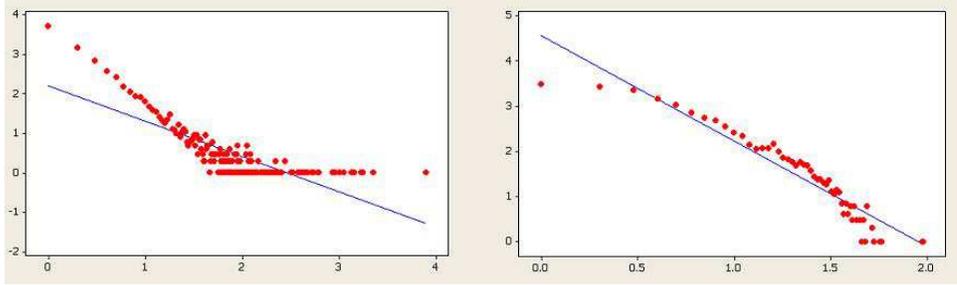}
\caption{Log-log scatterplot of $k_{in}$ and $k_{out}$ (respectively) for the Debian network}
\end{figure}
\section{Conclusion and Discussion}
This research has shown that package dependency networks mined from two open-source software repositories share the 
following properties typical to other real-world networks:
\begin{itemize}
\item The small-world effect: short geodesic path lengths and high clustering.
\item Near power-law distribution of edges.
\item The presence of a giant component, $|V \in \Omega_1| \gg |V \in \Omega_2|$
\end{itemize}
\par There are many directions for future research in the study of software networks. Currently, there is no model 
of network formation that takes software dynamics (reuse, refactoring, addition of new packages) in to account. Also, 
the impact of the network structure on software dynamics should be investigated. Future research should identify 
other networks in software and move towards formulating a theory of networks and their value to software engineering. Additional dependency networks can be constructed on Windows computers using 
memory profiling tools, and determining interactions based on shared .DLL (Dynamic Library Link) files and Active-X 
controls. 



\end{document}